\documentclass[aps,prl,reprint,superscriptaddress]{revtex4-2}

\usepackage[]{graphicx}
\usepackage[dvipsnames]{xcolor}

\makeatletter
\newcommand\footnoteref[1]{\protected@xdef\@thefnmark{\ref{#1}}\@footnotemark}
\makeatother
\begin{document}

%\preprint{}

\title{Transport properties of band engineered p-n heterostructures of epitaxial Bi$_2$Se$_3$/(Bi$_{1-x}$Sb$_x$)$_2$(Te$_{1-y}$Se$_y$)$_3$ topological insulators}


%T. Mayer, H. Werner, F. Schmid, R. Diaz-Pardo, I. Vobornik, J. Fujii, C. H. Back, M. Kronseder and D. Bougeard
\author{T. Mayer}
%\email[]{thomas.mayer@ur.de}
%\thanks{}
\affiliation{Institut f{\"u}r Experimentelle und Angewandte Physik, Universit{\"a}t Regensburg, D-93040 Regensburg, Germany}

\author{H. Werner}
%\email[]{}
%\thanks{}
\affiliation{Institut f{\"u}r Experimentelle und Angewandte Physik, Universit{\"a}t Regensburg, D-93040 Regensburg, Germany}

\author{F. Schmid}
%\email[]{}
%\thanks{}
\affiliation{Institut f{\"u}r Experimentelle und Angewandte Physik, Universit{\"a}t Regensburg, D-93040 Regensburg, Germany}

\author{R. Diaz-Pardo}
%\email[]{}
%\thanks{}
\affiliation{Fakult{\"a}t f{\"u}r Physik, Technische Universit{\"a}t M{\"u}nchen, D-85748 Garching b. M{\"u}nchen, Germany}

\author{J. Fujii}
%\email[]{}
%\thanks{}
\affiliation{Istituto Officina dei Materiali (IOM)-CNR, Laboratorio TASC, Area Science Park, S.S.14, Km 163.5, 34149 Trieste, Italy}

\author{I. Vobornik}
%\email[]{}
%\thanks{}
\affiliation{Istituto Officina dei Materiali (IOM)-CNR, Laboratorio TASC, Area Science Park, S.S.14, Km 163.5, 34149 Trieste, Italy}

\author{C. H. Back}
%\email[]{}
%\thanks{}
\affiliation{Fakult{\"a}t f{\"u}r Physik, Technische Universit{\"a}t M{\"u}nchen, D-85748 Garching b. M{\"u}nchen, Germany}

\author{M. Kronseder}
%\email[]{}
%\thanks{}
\affiliation{Institut f{\"u}r Experimentelle und Angewandte Physik, Universit{\"a}t Regensburg, D-93040 Regensburg, Germany}

\author{D. Bougeard}
\email[]{dominique.bougeard@ur.de}
%\thanks{}
\affiliation{Institut f{\"u}r Experimentelle und Angewandte Physik, Universit{\"a}t Regensburg, D-93040 Regensburg, Germany}


\date{\today}

\begin{abstract}
The challenge of parasitic bulk doping in Bi-based 3D topological insulator materials is still omnipresent, especially when preparing samples by molecular beam epitaxy (MBE). Here, we present a heterostructure approach for epitaxial BSTS growth. A thin n-type Bi$_2$Se$_3$ (BS) layer is used as an epitaxial and electrostatic seed which drastically improves the crystalline and electronic quality and reproducibility of the sample properties. In heterostructures of BS with p-type (Bi$_{1-x}$Sb$_x$)$_2$(Te$_{1-y}$Se$_y$)$_3$ (BSTS) we demonstrate intrinsic band bending effects to tune the electronic properties solely by adjusting the thickness of the respective layer. The analysis of weak anti-localization features in the magnetoconductance indicates a separation of top and bottom conduction layers with increasing BSTS thickness. By temperature- and gate-dependent transport measurements, we show that the thin BS seed layer can be completely depleted within the heterostructure and demonstrate electrostatic tuning of the bands via a back-gate throughout the whole sample thickness.
\end{abstract}

%\keywords{}

\maketitle

\section{I. Introduction}
Three-dimensional topological insulators (3D TIs), are predicted to feature helical topological surface states (TSS) with linear dispersion and time reversal symmetry protection \cite{PhysRevB.76.045302,PhysRevB.75.121306,PhysRevB.79.195322,RevModPhys.82.3045}. Experimentally, the first 3D TI was realized in Bi$_{1-x}$Sb$_x$ \cite{Hsieh2008}, sparking a vast amount of research especially around a whole family of mostly Bismuth-based compounds. The alloy Bi$_2$Se$_3$ (BS) was quickly identified as a promising member of this family. However, while ab-initio calculations showed a prototypical TI band structure \cite{Zhang2009}, angle-resolved photoemission spectroscopy (ARPES) measurements consistently revealed the Fermi energy (E$_\mathrm{F}$) to lie in the bulk conduction band because of donor-type Selenium vacancies and/or Se$_\mathrm{Bi}$ anti-sites \cite{Xia2009,doi:10.1002/adma.201200187}.
Electronic transport experiments are therefore often dominated by bulk states, making the full utilization of the unique TSS characteristics challenging. The presence of parasitic bulk conduction is generally shared by all Bi-based compounds and the strategies to counteract this issue have been multifold. Successful compensation of unintentional dopants has for example been achieved in single crystalline Bi-Sb-Te-Se solid solutions grown by the Bridgman technique, resulting in suppressed bulk conduction and surface dominated transport \cite{Xu2014}.\\
Next to the Bridgman method a widely spread approach to grow crystalline 3D TI samples is molecular beam epitaxy (MBE) that provides crucial advantages for many experimental and possible technological applications. For example, MBE offers quick adjustment of alloy stoichiometries, precise control of sample thickness down to single layers and the capability of in-situ preparation of hybrid devices with well defined interfaces, all while possibly opening a way to wafer-size scalability. However, sample quality has lacked behind significantly compared to other preparation methods and while the issue of parasitic bulk conduction due to structural disorder has not been conclusively solved, research especially concerning the promising quaternary alloy (Bi$_{1-x}$Sb$_x$)$_2$(Te$_{1-y}$Se$_y$)$_3$ (BSTS) has stalled.\\
In this contribution, we investigate MBE-grown BS/BSTS heterostructures within a vertical p-n-type concept. We show that BS acts as an excellent seed layer for epitaxial BSTS preparation already reducing unintentional doping due to improved crystallinity. Furthermore, we deliberately tune BSTS into a slight p-type regime via its stoichiometry and use the intrinsically n-type BS to create a band bending within the heterostructure by compensation of opposite excess charges.
In a systematic study, we investigate the transport properties of such heterostructures grown on SrTiO$_3$ (STO) and provide a recipe for a highly reproducible growth of BS/BSTS with minimized bulk conduction as-grown. Depending on the respective BS and BSTS thickness, we observe a strong suppression of trivial bulk conduction of the BS layer and a separation of the topological surface states. The choice of highly dielectric STO furthermore allows us to tune the electronic properties of the samples via back-gating leaving the top surface unoccupied for potential surface experiments or interfacing in hybrid devices.

\section{II. Experimental}
All samples presented in this work were grown by molecular beam epitaxy. The thicknesses of the layers were determined through reflective high energy electron diffraction (RHEED) oscillations. ARPES characterization was performed at 77$\,$K with a spot-size of 150$\,\mu$m$\,\times\,$50$\,\mu$m and a photo energy of 36$\,$eV for maximum photoemission intensity of the surface states with respect to bulk states. The ARPES samples were protected from oxidation by removable selenium capping layers. The samples for magnetotransport measurements were capped in-situ by 7$\,$nm Al$_2$O$_3$. All magnetotransport measurements were carried out at 4.2$\,$K, utilizing a standard 4-point, low-frequency lock-in technique in a Hall bar geometry with the magnetic field applied perpendicular to the film. The Hall bar has a width of $w = 20\, \mu$m and a length of $l = 300\, \mu$m. The obtained sheet resistance is defined as $R_\mathrm{S} = \frac{U_{\mathrm{xx}}}{I}\cdot \frac{w}{l}$,  where $I$ is the applied current and $U_{\mathrm{xx}}$ the measured longitudinal voltage. For electrostatic back-gating a voltage was applied between the sample and the bottom of the chip carrier with the STO substrate acting as a dielectric barrier. For additional front-gating the samples were covered by an insulating bilayer of 30$\,$nm SiO$_2$ and 100$\,$nm of Al$_2$O$_3$ and a gold electrode.

\section{III. Results}
\subsection{A. BS as seed layer}

\begin{figure}
	%\centering
	\includegraphics[scale=0.67]{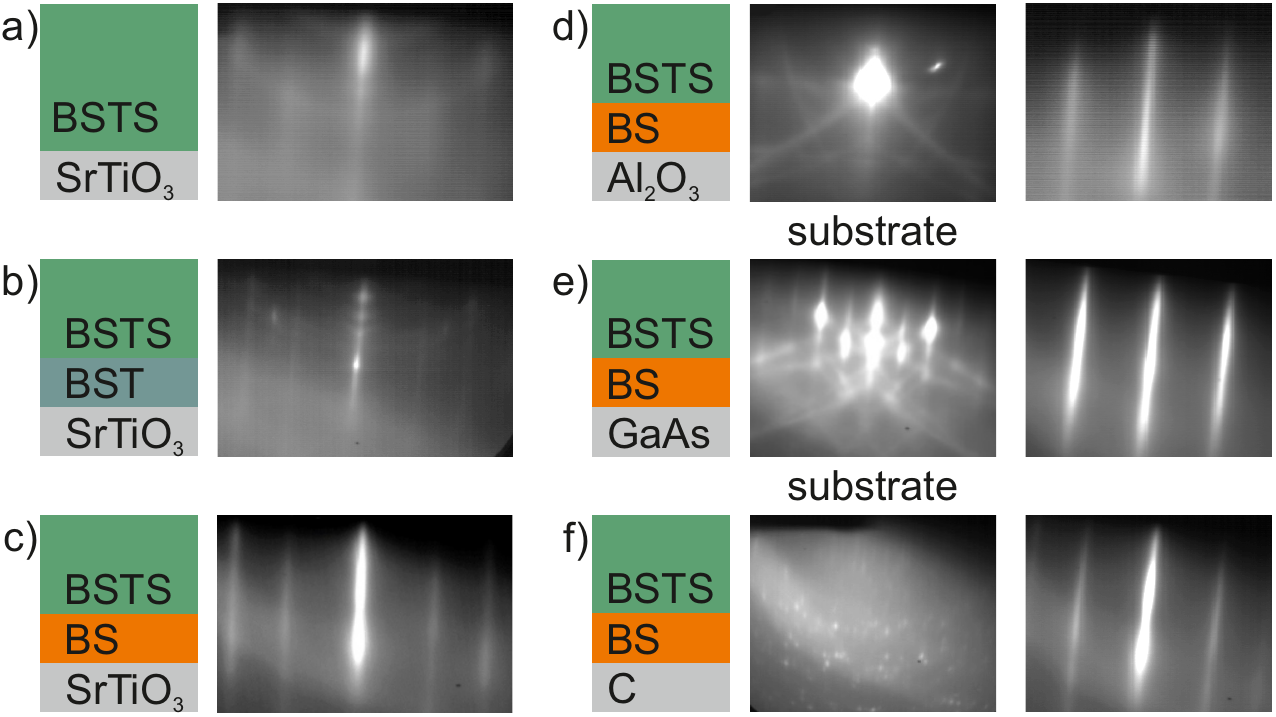}
	\caption{\label{}RHEED patterns of BSTS directly grown on STO (a), with BST seed layer (b) and BS seed layer (c). d) - f)  RHEED patterns of BSTS (right) with BS seed layer on different substrates (middle). }
	\label{Fig1}
\end{figure}
For Bi-based TI alloys, the optimization of growth quality is crucial since the electronic properties are largely governed by unintentional doping caused by lattice defects. The most widely investigated MBE-grown 3DTI is Bi$_2$Se$_3$. Its tetradymite crystal structure is built up by quintuple layers (1$\,$QL $\approx$ 1$\,$nm), with weak van der Waals (vdW) interlayer bonding between QLs, enabling successful growth on a variety of substrates via vdW epitaxy and high crystallinity was achieved by precise optimization of growth conditions \cite{Ginley_2016,doi:10.1002/pssb.202000007}. MBE of related ternary (e.g. Bi$_2$(Se$_{1-x}$Te$_x$)$_3$ or (Bi$_x$Sb$_{1-x}$)$_2$Te$_3$) and especially quaternary compounds like BSTS has been less intensively studied. Expanding the alloy complicates the growth procedure, it increases the amount of atomic disorder naturally occurring in those systems and reduces the amount of suitable substrates since a large lattice mismatch induces crystal defects. \\
Investigating epitaxial preparation of BSTS directly on the STO(111) substrate, we were unable to find a reliable and reproducible regime of sole single-crystalline order and routinely observed patterns with poly-crystalline features in RHEED imaging during growth, as examplarily shown in Fig. \ref{Fig1}a) for 6$\,$QL BSTS with $(x|y)=(70|90)$. Using a BST seed layer (Fig. \ref{Fig1}b) lead to improvements, but caused 3D features in the RHEED pattern. In addition, different substrates or BSTS stoichiometries compel an adaptation of growth parameters. Introducing a BS seed layer, however, facilitates the growth of high-quality BSTS films, independently of its stoichiometry and the used substrate. Surprisingly, even a single BS layer acts as a highly oriented vdW seed and is sufficient to ease the vdW epitaxy of subsequent BSTS. The protocol to grow the BS seed layers is as follows: saturating the substrate surface by Se at 190$^\circ\,$C  for 150$\,$s, growing the BS layers while ramping the substrate temperature from 190$^\circ\,$C to 250$^\circ\,$C within the first 2$\,$QL, followed by annealing under constant Se-flux at 290$^\circ\,$C. At the BSTS growth temperature of 255$^\circ\,$C, the RHEED pattern shows pronounced oscillations and no indication of 3D or poly-crystalline features. Next to STO(111) (see Fig. \ref{Fig1}c), this protocol has successfully been applied to a variety of substrates, also beyond the common (111)-orientation \cite{doi:10.1002/pssb.202000007}, without requiring to change the growth protocol. Figures \ref{Fig1} c)-e) exemplarily show RHEED patterns of TI samples (right) and the respective substrate (middle), demonstrating single-crystalline growth on Al$_2$O$_3$(0001) (c), GaAs(111) (d), and even disordered C(111) (f). In addition, successful growth was achieved on Al$_2$O$_3$ (11-20), GaAs (001) and InP(111).\\
Crucial for the electronic properties, we found BS to also function as an "electrostatic seed" layer. Selenium vacancies and Se$_\mathrm{Bi}$ anti-sites lead to a large bulk donor level in BS, as shown in Fig. 2a) \cite{doi:10.1002/adma.201200187,Chen178,RUMANN2019258}. This pins the Fermi level to the bulk conduction band. It therefore reproducibly fixes the starting point for subsequent layers to an n-type foundation independent of the used substrate. BSTS growth directly on a substrate lead to strong variations of the samples' electronic properties even for constant stoichiometry. Since the interface potential between sample and substrate is susceptible to minor fluctuations of growth conditions, a controlled positioning of E$_\mathrm{F}$ in the band gap throughout the complete sample thickness has proven to be challenging.
Hence, the epitaxial and the electrostatic seed layer functionality of BS dramatically improved the quality, controllability, and especially the reproducibility of crystallographic and electronic sample properties, as we will demonstrate in the following. 

\subsection{B. Heterostructure concept}

\begin{figure*}
	\centering
	\includegraphics[width=0.8\textwidth]{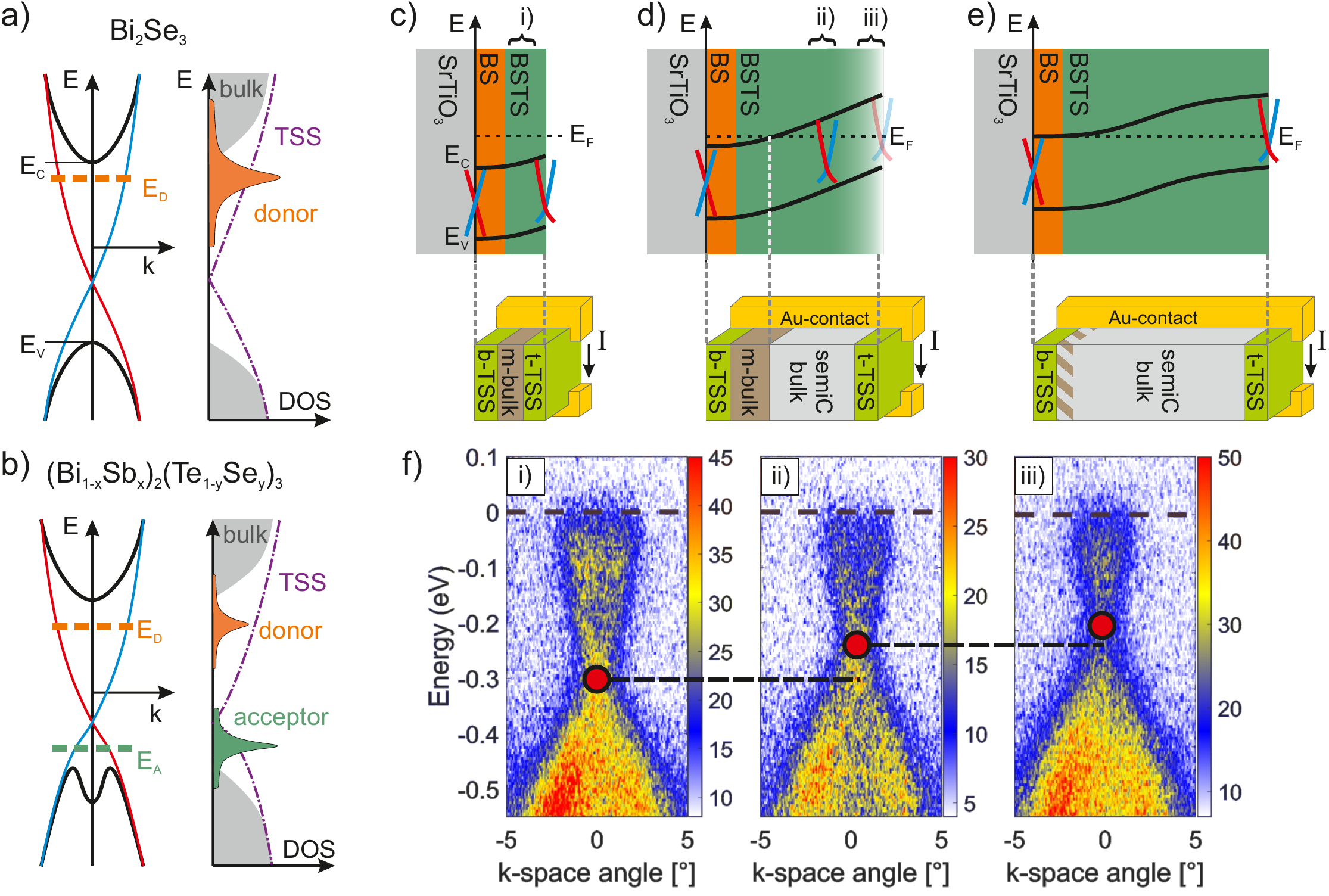}
	\caption{Schematic band structure and density of states (DOS) of dopant levels in BS (a) and BSTS (b). c) Heterostructure concept: depending on respective BS and BSTS thicknesses, band bending is introduced within the bilayer, leading to different sizes of metal-like (m-bulk) and semiconductor-like (semiC) bulk contribution additional to conduction of top (t-TSS) and bottom (b-TSS) topological surface states. f) ARPES at 77$\,$K imaging the band bending evolution for samples of 1 QL BS and 3 (i), 6 (ii) and 12 QL (iii) BSTS. The horizontal black dashed line represents the Fermi level. The black dashed triangles are a guide to the eye to the position of the t-TSS}
\end{figure*}
While the implementation of a BS seed layer proved to be highly favorable for epitaxial BSTS preparation, its prevalent bulk donor level potentially adds a large contribution to the overall bulk conductance of the bilayer. On the other hand, (Bi$_{1-x}$Sb$_x$)$_2$(Te$_{1-y}$Se$_y$)$_3$ allows an engineering of key band structure features, especially the fine tuning of the effective donor to acceptor ratio via the stoichiometric parameters $x$ and $y$ \cite{PhysRevB.84.165311,Arakane2012,PhysRevB.87.085442,PhysRevB.93.245149}. Based on a test series, we chose $x = 70-74\%$ and $y = 87-91\%$, aiming to maximize the BSTS band gap while creating a slight acceptor surplus (Fig. 2b). A heterostructure with the hence p-type BSTS and n-type BS generates a bending of the system's electronic bands \cite{doi:10.1002/pssr.201206391,Eschbach2015,PhysRevB.96.125125}, schematically pictured in Figs. 2c)-e).
For very thin BSTS, opposite excess charges begin to compensate, but the effect is too small and the Fermi level stays above the conduction band minimum (CBM) (Fig. 2c), as is revealed by ARPES on a heterostructure with 1$\,$QL BS and 3$\,$QL BSTS in Fig. 2f)i), where an occupation of the bulk conduction band can be observed. Increasing the BSTS thickness enhances the band bending until E$_\mathrm{F}$ is pulled below the CBM into the energy gap at the top surface (Fig. 2d ii). The color grading towards iii) in Fig. 2d) indicates the evolution of the shift when increasing the BSTS thickness. This behavior is verified by ARPES in Fig. 2f)ii) and iii). Ideally, at some point, the band bending is sufficient to pull E$_\mathrm{F}$ into the band gap almost throughout the whole heterostructure by completely depleting the BS layer (Fig 2e). It is important to stress that while our ARPES measurements follow the trend expected for this thickness-dependent band bending, they only image the energy bands at the very surface of the sample. Electrical transport properties, however, are governed by the complete band structure throughout the whole sample thickness. In the most general case, illustrated in Fig. 2d), the sample can be divided into three segments contributing to transport: a semiconductor-like channel (semiC bulk) where E$_\mathrm{F}$ lies in the band gap, a trivial, metal-like bulk channel (m-bulk) for E$_\mathrm{F}$ intersecting the conduction band and the non-trivial top and bottom TSS (t-TSS, b-TSS).

\subsection{C. Magnetotransport characterization}
\begin{figure}[b]
	\centering
	\includegraphics[scale=0.85]{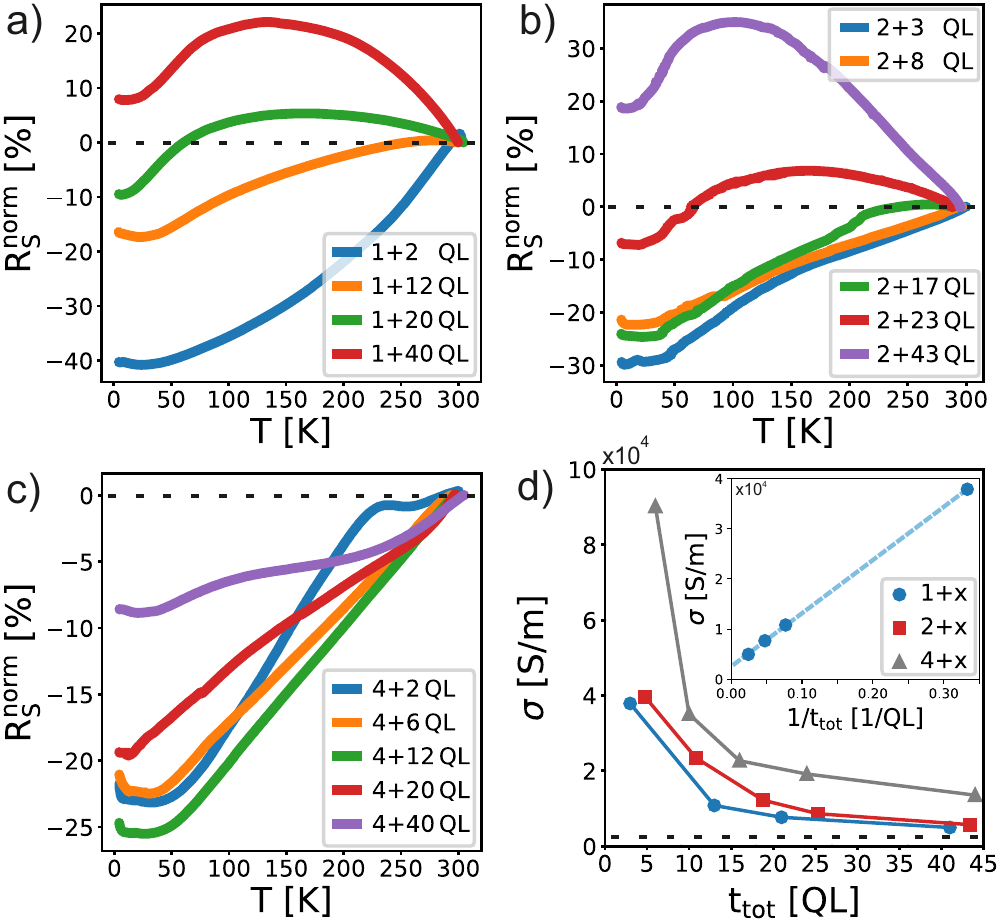}
	\caption{Sheet resistance as a function of temperature normalized to room temperature of 1+$x$ (a), 2+$x$ (b), and 4+$x$ (c) series. d) Conductivity at 4.2$\,$K of all three series versus total sample thickness. The inset shows the 1+$x$ series versus 1/$t_{\mathrm{tot}}$ and a linear fit (light blue dashed line). The y-intercept yields the asymptote (black dashed line) in the main figure. }
\end{figure}
\noindent

To study the contributions of these channels to electronic transport, systematic BSTS thickness series are investigated with 1, 2 and 4 QL of BS seed layers, in the following referred to as 1+$x$, 2+$x$ and 4+$x$ series, with the BSTS thickness $x$ reaching from 2 to 43$\,$QL. Figures 3a)-c) show the temperature dependence of the normalized sheet resistance $R_\mathrm{S}^\mathrm{norm}$($T$) = $R_\mathrm{S}$($T$)/$R_\mathrm{S}$(300$\,$K) - 1 for the three series. The different transport contributions manifest in the measurements due to their different temperature dependencies. For the semiconductor-like channel, activated carriers freeze out upon reducing the temperature and  $R_\mathrm{S}$ increases. The trivial metal-like bulk conduction and the TSS, on the other hand, act like metals: $R_\mathrm{S}$ decreases towards lower temperatures due to the reduction of electron-phonon scattering \cite{Gao2012}. The competition of these three transport channels as a function of BSTS thickness is observed for the 1+$x$ (Fig. 3a) and 2+$x$ (Fig. 3b) series. Similar to many observations in bulk conducting TIs, the thinnest samples show a strict metallic behavior due to E$_\mathrm{F}$ lying above the conduction band edge, corresponding to Fig. 2c). As expected from the sketch of Fig. 2d), this trivial metallic contribution gradually diminishes with growing BSTS thickness, leading to the semiconductor-like contribution beginning to dominate $R_\mathrm{S}$(T) at high temperatures. For the thickest samples of Figs. 3a) and b) $R_\mathrm{S}$ increases to about 120$\,$K, before a small metallic decrease is observed. This behavior has been reported for fully bulk compensated TIs. There, the drop of $R_\mathrm{S}$ at low temperatures is ascribed to dominant TSS transport \cite{Xu2014,Arakane2012}. A stark contrast is presented by the $R_\mathrm{S}$(T) behavior of the 4+$x$ series in Fig. 3c). Here, the thicker BS layer leads to a trivial metal-like bulk channel large enough to dominate transport for all BSTS thicknesses in the complete temperature range.\\ 
These observations are confirmed by plotting the conductivity $\sigma$ at 4.2$\,$K versus the total sample thickness $t_\mathrm{tot}$ (Fig. 3d, see supplemental material for resistivity values \cite{supplemental}). All three series show a significant decrease of $\sigma$ with increasing $t_\mathrm{tot}$. The dashed line is obtained from the y-intercept of the linear fit of the 1+$x$ series in the 1/$t_{\mathrm{tot}}$ depiction (inset) and therefore represents the bulk conductivity of BSTS in the limit of $t\rightarrow \infty$ \footnote{Within each series, the BS thickness remains constant while the BSTS thickness gradually increases. The limit of $t\rightarrow \infty$ therefore means that the BSTS thickness approaches infinity: any thickness independent contribution to conduction like the TSS or any contribution from the BS vanishes.}. We find a comparatively low value of $\approx$2500$\,$S/m. This non-zero bulk conductivity is commonly ascribed to randomly distributed charge puddles and thermally activated carriers from acceptor and donor levels \cite{PhysRevLett.109.176801,Skinner2013,PhysRevB.87.165119,PhysRevB.93.245149,Rischau_2016}. Any offset from the dashed line is expected to mainly stem from a trivial bulk contribution caused by the BS seed layer or TSS conduction. The 1+x series (blue circles) approaches the asymptote slightly more quickly than the 2+$x$ samples (red squares), but for thicknesses above $\sim$20$\,$QL both curves begin to converge. Again, the 4+$x$ series (grey triangles) provides a contrast in showing a significantly larger conductivity for all thicknesses. These observations confirm the conclusions already drawn from the $R_\mathrm{S}$(T) measurements: Using 4$\,$QL of BS induces a large metal-like bulk channel. It dominates the $R_\mathrm{S}$(T) behavior and also substantially contributes to the overall conductivity of all samples at 4.2$\,$K. With 1$\,$QL or 2$\,$QL a qualitatively different behavior is observed: For sufficient BSTS thickness ($>20\,$QL) the BS contribution seems to become negligible, yielding almost identical conductivity very close to the value of bulk BSTS. This indicates that we have approached the ideal case of Fig. 2e).

\begin{figure}
	\centering
	\includegraphics[scale=0.85]{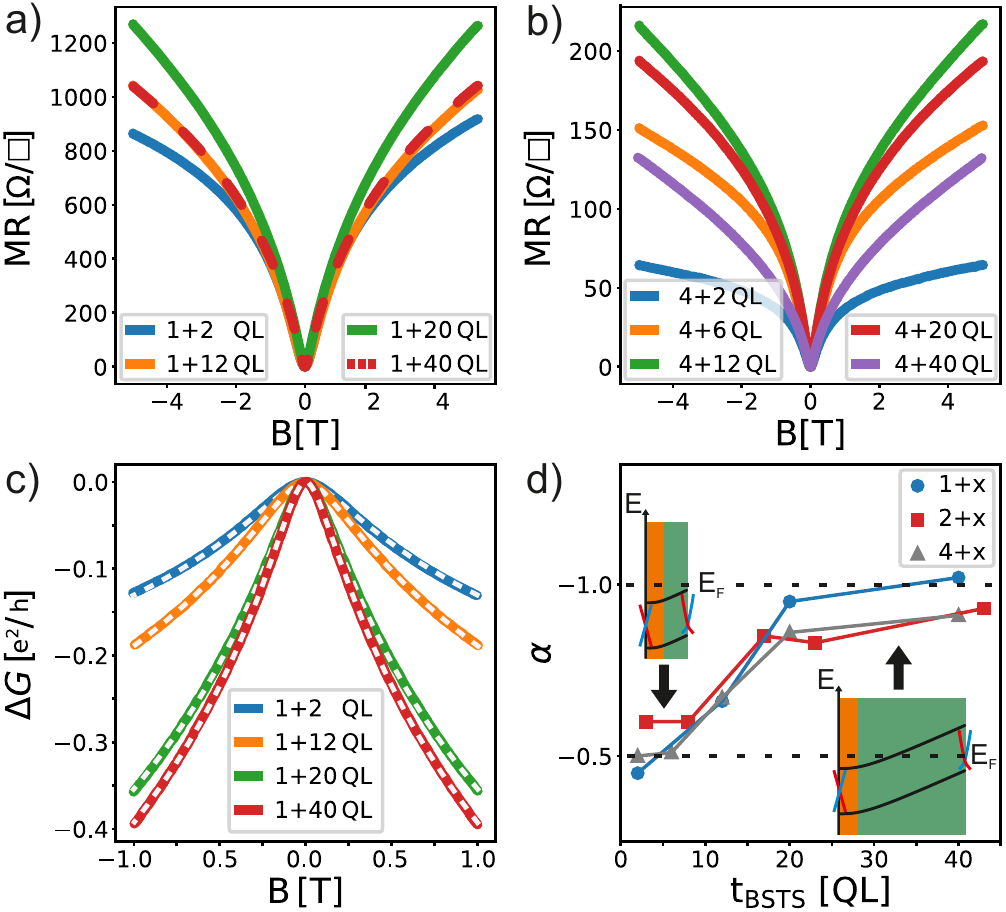}
	\caption{Magnetoresistance at 4.2$\,$K for the 1+$x$ (a) and 4+$x$ (b) series. c) HLN fits (white dotted lines) to $\Delta G(B)$ for the 1+x series. d) $\alpha$ values from HLN fits for all three series versus BSTS thickness at 4.2$\,$K. The insets show band structure sketches for different BSTS thicknesses corresponding to the evolution of $\alpha$.}
\end{figure}
To further characterize the samples, we applied a perpendicular magnetic field B. We find that the significant contribution of the m-bulk channel in the 4+x series is confirmed in the Hall resistance (see supplemental material \cite{supplemental}). Figures 4a) and b) compare the absolute magnetoresistance $MR\,$(B) =  $R_\mathrm{S}$(B) - $R_\mathrm{S}$(0$\,$T) at 4.2$\,$K of the 1+$x$ and 4+$x$ series. In all measurements a characteristic cusp-like, positive MR around zero magnetic field is observed, commonly described to stem from weak anti-localization \cite{10.1117/12.2063426}. Transport mediated by TSS in TIs is expected to be especially sensitive to this effect, due to their spin helicity and the arising $\pi$ Berry phase \cite{10.1117/12.2063426, PhysRevB.86.035422}. Applying a perpendicular magnetic field breaks time-reversal symmetry and therefore lifts the enhanced delocalization, causing an increase of sample resistance with magnetic field. We conclude on the TSS to be the main source of MR, reasoning by exclusion between the three relevant transport-contributing channels introduced in Fig. 2 c)-e) (metallic bulk channel, semiconductor bulk channel, TSS). Comparing first the 1+x (small metallic bulk contribution) and the 4+x series (larger metallic bulk contribution): if the metallic bulk channel were the dominant source of the observed MR, the absolute MR values for the 4+x series should be larger than for 1+x. This is not the case, since the MR values for the 4+x series is significantly smaller than for the 1+x. As a consequence, we exclude the metallic bulk contribution to be dominantly responsible for the observed MR. Considering the semiconductor contribution, we have previously discussed that it will increase with increasing BSTS thickness. In our data, however, the MR of 1+40 is identical to 1+12 and smaller than the one of 1+20. This excludes the semiconducting channel to be responsible for the observed MR behavior. We thus interpret the observed MR behavior as a manifestation of the presence of the TSS. In addition to the cusp-signature around zero field, a transition to quadratic or linear behavior at higher magnetic fields is often reported in TIs. Whereas the quadratic behavior is widely accepted to stem from 3D bulk conduction \cite{Gao2012}, the linear MR is subject to more discussion. In the measurements of Figs. 4a) and b) we never observe signatures $\sim$B$^2$, again suggesting the absence of a sizeable 3D bulk contribution in all our samples. In the 4+$x$ series (Fig. 4b) the MR approaches a linear regime above $\approx$2.5T. In contrast, for 1$\,$QL seed layer (Fig. 4a) the cusp-behavior prevails in the complete investigated field range. 
To explain the origin of such a linear MR (LMR) mainly two models have been proposed: The quantum model of Abrikosov yields an LMR as consequence of linear dispersion, which could link the observation to the linearly dispersive surface states of TIs \cite{PhysRevB.58.2788}. The classical model by Parish and Littlewood, however, shows an LMR to emerge in inhomogeneous two dimensional trivial conductors \cite{Parish2003,PhysRevB.72.094417}. Since we only observe an LMR in the 4+x series, we conclude the classical model to be a more likely explanation. As evaluated above, 4$\,$QL of BS seed layer lead to a significant, but very thin, metal-like bulk conduction channel largely dominating transport. In the framework of Parish and Littlewood, this channel could be subject to LMR that superimposes with the cusp-like WAL behavior of the TSS. This metallic bulk contribution could furthermore serve as an explanation for the strikingly smaller overall magnetoresistances observed in the 4+x series. It acts as a channel parallel to the TSS and therefore reduces the ratio of transport channels underlying weak anti-localization effects.\\
For a more detailed analysis of the WAL signature observed in TIs, the theory of Hikami, Larkin and Nagaoka (HLN) \cite{10.1143/PTP.63.707} is commonly applied in the literature to fit the measured data with
\[
\Delta G_{\mathrm{HLN}}(B) = \alpha \frac{\mathrm{e}^2}{\mathrm{\pi} \mathrm{h}} \left[ \Psi \left( \frac{\mathrm{\hbar}}{4\mathrm{e}Bl_\mathrm{\phi}^2} + \frac{1}{2} \right) - \ln \left(\frac{\mathrm{\hbar}}{4\mathrm{e}Bl_\mathrm{\phi}^2}\right)\right],
\]
assuming $G(B) \approx 1/R_\mathrm{S} (B)$ and $\Delta G_{\mathrm{HLN}}(B) \approx$ $\Delta G(B)  \equiv$ $G(B)$ - $G$(0). Since $\Delta G(B)$ is directly obtained from the magnetoresistance measurements presented in Fig. 4a), it is important to note that it is, in general, not free from bulk contributions that may not underlie weak anti-localization. The origins and signatures of such additional contributions to MR, as well as their influence on HLN accuracy, need to be subject to a more thorough investigation. In the above equation, e is the elementary charge, h Planck's constant and $\psi$ the digamma function. The free fit parameters are the phase coherence length $l_\Phi$ and the dimensionless pre-factor $\alpha$. The simplectic case of the HLN theory, distinguished by strong spin-orbit coupling and the absence of magnetic scattering, is usually associated to topological insulators \cite{PhysRevB.86.035422}. It is expected to yield a value of $\alpha = -0.5$ per independent parallel channel contributing to conduction. Figure 4c) exemplarily shows the HLN fits (white dashed line) to $\Delta G$ in the 1+$x$ series, demonstrating a very good agreement of the theory and the measured data within $\pm$1$\,$T. The $\alpha$-values for all three series obtained from this fit interval are plotted as a function of the respective BSTS thickness $t_{\mathrm{BSTS}}$ in Fig. 4d) and a striking resemblance, independent of the seed layer, is observed. For small $t_{\mathrm{BSTS}}$, $\alpha$ starts around a value of -0.5, before an increase sets in, approaching -1 above 20$\,$QL. For the smallest BSTS thickness we expect the Fermi level to lie above the conduction band edge throughout the complete heterostructure (see upper inset). Hence, the whole sample effectively acts as one conducting channel and $\alpha = -0.5$ is expected. It has furthermore been suggested that below a thickness of approximately 10$\,$QL, $\alpha = -0.5$ would even be expected for separated channels due to coupling of top and bottom TSS mediated by tunneling or hopping \cite{PhysRevLett.113.026801,Wang2016}. This could explain the simultaneous increase in all three series to start around this threshold. The approach of $\alpha = -1$  above 20$\,$QL then suggests a true separation of two independent conduction channels. Contrary to a common interpretation, our data shows that $\alpha = -1$ not necessarily allows to conclude a completely insulating bulk, where the TSS at top and bottom surface each contribute -0.5 to $\alpha$. We have shown that the 4+$x$ series clearly shows significant bulk conduction for all BSTS thicknesses. However, our analysis indicates that with increasing t$_{\mathrm{BSTS}}$ the upper TSS still decouples from this bulk channel regardless of seed layer thickness. The lower inset of Fig. 4d) illustrates this more general case with the bottom TSS in direct contact with bulk states and a separated top TSS. In Ref. \cite{PhysRevB.86.035422} it has been theoretically predicted that this configuration can also yield $\alpha = -1$.\\
\begin{figure}
	\centering
	\includegraphics[scale=0.85]{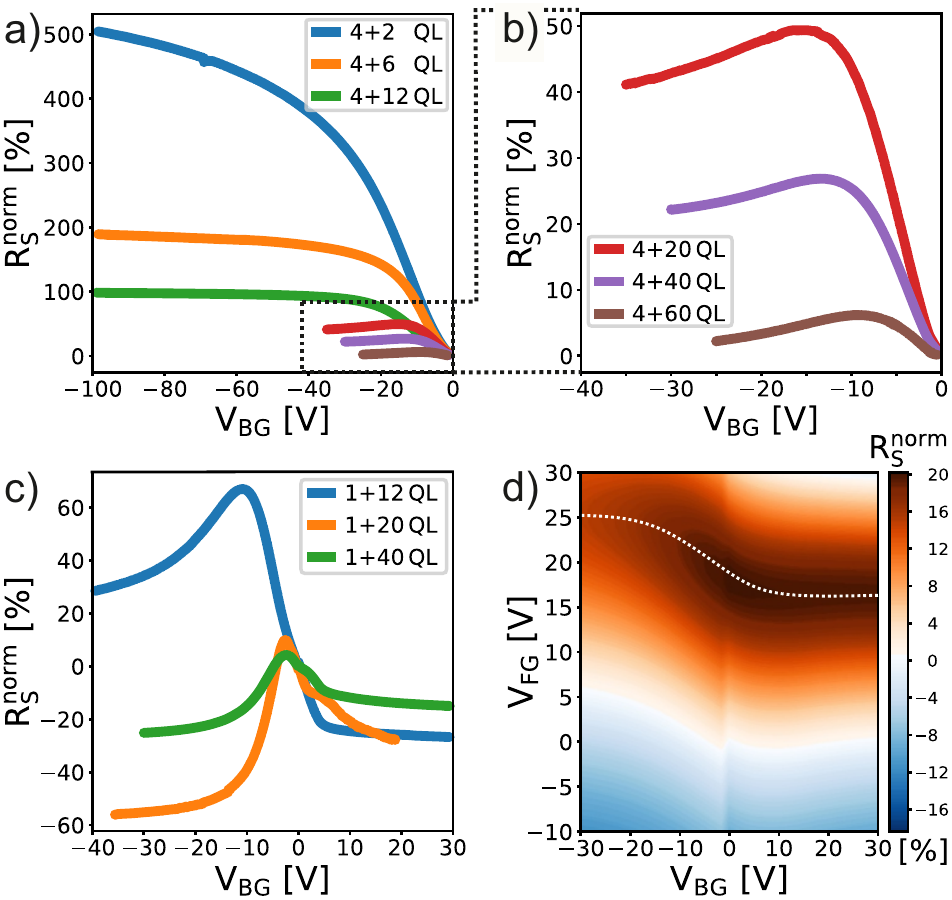}
	\caption{Normalized sheet resistance against back-gate voltage at 4.2$\,$K for the 4+$x$ series (a) with a zoom-in for the thickest samples (b) and the 1+$x$ series (c). d) Dual-gated measurement of sheet resistance for sample 1+40. The white dashed line is a guide to the eye along the maximum of $R_\mathrm{S}^\mathrm{norm}$.}
\end{figure}
In addition to optimization of as-grown electronic properties, the heterostructure approach using a BS seed layer enables direct epitaxial growth on SrTiO$_3$ (STO). Due to its ultrahigh relative permittivity at cryogenic temperatures \cite{PhysRev.174.613,PhysRevB.19.3593}, STO can function as back-gate (BG) dielectric allowing easy implementation of electrostatic gating to further investigate and modify transport behavior by means of a field effect-induced shift of the chemical potential. Figures 5 a)-c) compare the relative change of $R_\mathrm{S}$ versus back-gate voltage V$_\mathrm{BG}$ defined as $R_\mathrm{S}^\mathrm{norm}$(V$_\mathrm{BG}$) = $R_\mathrm{S}$(V$_\mathrm{BG}$)/$R_\mathrm{S}$(0$\,$V) - 1 of 4+$x$ (Fig. 5a and b) and 1+$x$ (Fig. 5b) samples at 4.2$\,$K. Applying a negative voltage causes an upward band bending. Samples 4+2, 4+6, and 4+12 (Fig. 5a) therefore show a steep increase of $R_\mathrm{S}^\mathrm{norm}$ for $V_\mathrm{BG}<0$ since the predominant n-type metallic bulk conduction in theses samples is reduced by the induced field effect. $R_\mathrm{S}^\mathrm{norm}$(V$_\mathrm{BG}$) reaches values of 500\%, 200\%, and 100\%, respectively, but no maximum is observed as would be expected when tuning the chemical potential through the charge neutrality point (CNP). We ascribe this to the effective screening of the BG-induced electric field by the large metal-like conduction channel at the STO/TI interface before the CNP is reached. Intriguingly, starting from a BSTS thickness of 20$\,$QL the behavior changes and a maximum of $R_\mathrm{S}^\mathrm{norm}$ with respect to the voltage is observed that shifts towards 0$\,$V with increasing BSTS thickness (Fig. 5b). We interpret this observation as direct evidence for the compensation of opposite excess charges within the p-n-heterostructure: for thin BSTS, the metal-like bulk conduction channel at the STO/TI interface induced by the 4$\,$QL of strongly n-type BS remains large enough to screen the static electric field. Increasing the BSTS thickness, however, can sufficiently deplete the BS layer and BG-tunability is achieved. 
Figure 5c) shows $R_\mathrm{S}^\mathrm{norm}$($V_\mathrm{BG}$) for the 1+$x$ series. From the measurements of Fig. 3, we conclude that sample 1+12 still shows a small remaining metal-like bulk channel as-grown that practically vanishes for 20$\,$QL and 40$\,$QL of BSTS. These observations directly manifest in the BG behavior of the respective sample: Applying a positive voltage yields a downward bending of the energy bands that hence increases the remaining n-type bulk channel in sample 1+12. For +30$\,$V$\geq$ $V_\mathrm{BG}$ $\gtrsim$ +5$\,$V this channel is then large enough to screen the field effect, equivalent to the observations in thin 4+$x$ samples. Thus, $R_\mathrm{S}^\mathrm{norm}$ remains constant. Below 5$\,$V the BG tunability is regained and $R_\mathrm{S}^\mathrm{norm}$ increases up to a pronounced maximum at -10.5$\,$V. In stark contrast, samples 1+20 and 1+40 only show a small increase with a maximum at $V_\mathrm{BG}\approx-2\,$V. This clearly indicates the as-grown depletion of the metal-like bulk conduction channel induced by the BS seed layer. As observed for the other samples, any remaining contribution of this channel to electrical transport would be largely tunable by the field effect and therefore cause a significant increase of $R_\mathrm{S}$ for negative voltages in comparison to the unbiased measurement.\\
To gain deeper insight into the capability of BG-induced $R_\mathrm{S}$ tuning, Figure 5d) shows a dual-gate \footnote{To account for hysteresis effects, the front-gate was swept at 4.2$\,$K from +30$\,$V to -30$\,$V twice before the measurement was started. Then, the back-gate was set in steps of 1.5$\,$V in positive direction. After 10 minutes, to prevent time-dependent influences, the front-gate was swept from +30$\,$V to -30$\,$V with 60$\,$mV/s. After the measurement, the sample was brought to room-temperature to reset both gates and the procedure was repeated with the back-gate being set into negative direction. Due to this required reset of the gates a small discrepancy between the measurements for positive and negative back-gate direction is unavoidable. To account for this, the $R_\mathrm{S}$ values for positive back-gate voltage were shifted by 3.3$\,$V front-gate in Fig. 5d).}  measurement of sample 1+40. The white dashed line follows the maximum of $R_\mathrm{S}$ with respect to both gate voltages. For +30$\,$V$\geq$ $V_\mathrm{BG}$ $\gtrsim$ +10$\,$V the maximum of $R_\mathrm{S}$ stays almost constant and remains at the same $V_\mathrm{FG}\approx$16.5$\,$V. In this regime, the large positive BG again leads to a large metal-like conduction channel screening the BG field effect. Between a BG voltage of 10$\,$V and 0$\,$V the global maximum of $R_\mathrm{S}$ is reached. Furthermore, in the same $V_\mathrm{BG}$ range, the location of the maximum in terms of FG voltage starts to shift to higher values. This shift shows a coupling between BG and FG field effects demonstrating the BG's capability to tune the sample's electronic properties throughout its complete thickness even for 40$\,$QL of BSTS.

\section{IV. Conclusion}
In this study, we have presented an approach to band structure engineering  of 3DTI thin films by means of epitaxial MBE growth. Introducing down to a single QL BS seed layer lead to a significant improvement of the BSTS growth quality drastically reducing structural disorder and therefore reducing unintentional bulk doping. By varying the respective thicknesses of the n-type BS and p-type BSTS we were able to substantially tune the as-grown electronic properties of the heterostructures and disentangle the different contributions to the electronic transport of the occurring channels by temperature-, magnetic field-, and gate-dependent measurements. We have shown that the p-n-type architecture of our samples leads to a compensation of opposite excess charges, culminating in a complete depletion of metal-like bulk conduction for 1$\,$QL BS seed layer and a BSTS thickness above 20$\,$QL. By applying the theoretical framework of Hikami, Larkin and Nagaoka \cite{10.1143/PTP.63.707}, we observed a gradual formation of two separated conduction channels with increasing BSTS thickness, independent of the seed layer thickness, revealing a decoupling of at least the top-TSS from bulk states. The chosen STO substrate allowed the application of back-gating that was shown to be capable of modifying the samples' electronic properties throughout the whole thickness. This tuning capability, without occupying the top surface, in combination with the decoupling of the top TSS, is particularly attractive for surface experiments or the implementation in hybrid devices.

\section{Acknowledgments}
\begin{acknowledgments}
We acknowledge the financial support of the Deutsche Forschungsgemeinschaft through project ID~422~314695032-SFB1277 (subproject A01). We thank Magdalena Marganska, Klaus Richter, Cosimo Gorini, and Michael Barth for fruitful discussions.
\end{acknowledgments}

\bibliography{references}

\begin{thebibliography}{38}%
\makeatletter
\providecommand \@ifxundefined [1]{%
 \@ifx{#1\undefined}
}%
\providecommand \@ifnum [1]{%
 \ifnum #1\expandafter \@firstoftwo
 \else \expandafter \@secondoftwo
 \fi
}%
\providecommand \@ifx [1]{%
 \ifx #1\expandafter \@firstoftwo
 \else \expandafter \@secondoftwo
 \fi
}%
\providecommand \natexlab [1]{#1}%
\providecommand \enquote  [1]{``#1''}%
\providecommand \bibnamefont  [1]{#1}%
\providecommand \bibfnamefont [1]{#1}%
\providecommand \citenamefont [1]{#1}%
\providecommand \href@noop [0]{\@secondoftwo}%
\providecommand \href [0]{\begingroup \@sanitize@url \@href}%
\providecommand \@href[1]{\@@startlink{#1}\@@href}%
\providecommand \@@href[1]{\endgroup#1\@@endlink}%
\providecommand \@sanitize@url [0]{\catcode `\\12\catcode `\$12\catcode
  `\&12\catcode `\#12\catcode `\^12\catcode `\_12\catcode `\%12\relax}%
\providecommand \@@startlink[1]{}%
\providecommand \@@endlink[0]{}%
\providecommand \url  [0]{\begingroup\@sanitize@url \@url }%
\providecommand \@url [1]{\endgroup\@href {#1}{\urlprefix }}%
\providecommand \urlprefix  [0]{URL }%
\providecommand \Eprint [0]{\href }%
\providecommand \doibase [0]{https://doi.org/}%
\providecommand \selectlanguage [0]{\@gobble}%
\providecommand \bibinfo  [0]{\@secondoftwo}%
\providecommand \bibfield  [0]{\@secondoftwo}%
\providecommand \translation [1]{[#1]}%
\providecommand \BibitemOpen [0]{}%
\providecommand \bibitemStop [0]{}%
\providecommand \bibitemNoStop [0]{.\EOS\space}%
\providecommand \EOS [0]{\spacefactor3000\relax}%
\providecommand \BibitemShut  [1]{\csname bibitem#1\endcsname}%
\let\auto@bib@innerbib\@empty
%</preamble>
\bibitem [{\citenamefont {Fu}\ and\ \citenamefont
  {Kane}(2007)}]{PhysRevB.76.045302}%
  \BibitemOpen
  \bibfield  {author} {\bibinfo {author} {\bibfnamefont {L.}~\bibnamefont
  {Fu}}\ and\ \bibinfo {author} {\bibfnamefont {C.~L.}\ \bibnamefont {Kane}},\
  }\href {https://doi.org/10.1103/PhysRevB.76.045302} {\bibfield  {journal}
  {\bibinfo  {journal} {Phys. Rev. B}\ }\textbf {\bibinfo {volume} {76}},\
  \bibinfo {pages} {045302} (\bibinfo {year} {2007})}\BibitemShut {NoStop}%
\bibitem [{\citenamefont {Moore}\ and\ \citenamefont
  {Balents}(2007)}]{PhysRevB.75.121306}%
  \BibitemOpen
  \bibfield  {author} {\bibinfo {author} {\bibfnamefont {J.~E.}\ \bibnamefont
  {Moore}}\ and\ \bibinfo {author} {\bibfnamefont {L.}~\bibnamefont
  {Balents}},\ }\href {https://doi.org/10.1103/PhysRevB.75.121306} {\bibfield
  {journal} {\bibinfo  {journal} {Phys. Rev. B}\ }\textbf {\bibinfo {volume}
  {75}},\ \bibinfo {pages} {121306} (\bibinfo {year} {2007})}\BibitemShut
  {NoStop}%
\bibitem [{\citenamefont {Roy}(2009)}]{PhysRevB.79.195322}%
  \BibitemOpen
  \bibfield  {author} {\bibinfo {author} {\bibfnamefont {R.}~\bibnamefont
  {Roy}},\ }\href {https://doi.org/10.1103/PhysRevB.79.195322} {\bibfield
  {journal} {\bibinfo  {journal} {Phys. Rev. B}\ }\textbf {\bibinfo {volume}
  {79}},\ \bibinfo {pages} {195322} (\bibinfo {year} {2009})}\BibitemShut
  {NoStop}%
\bibitem [{\citenamefont {Hasan}\ and\ \citenamefont
  {Kane}(2010)}]{RevModPhys.82.3045}%
  \BibitemOpen
  \bibfield  {author} {\bibinfo {author} {\bibfnamefont {M.~Z.}\ \bibnamefont
  {Hasan}}\ and\ \bibinfo {author} {\bibfnamefont {C.~L.}\ \bibnamefont
  {Kane}},\ }\href {https://doi.org/10.1103/RevModPhys.82.3045} {\bibfield
  {journal} {\bibinfo  {journal} {Rev. Mod. Phys.}\ }\textbf {\bibinfo {volume}
  {82}},\ \bibinfo {pages} {3045} (\bibinfo {year} {2010})}\BibitemShut
  {NoStop}%
\bibitem [{\citenamefont {Hsieh}\ \emph {et~al.}(2008)\citenamefont {Hsieh},
  \citenamefont {Qian}, \citenamefont {Wray}, \citenamefont {Xia},
  \citenamefont {Hor}, \citenamefont {Cava},\ and\ \citenamefont
  {Hasan}}]{Hsieh2008}%
  \BibitemOpen
  \bibfield  {author} {\bibinfo {author} {\bibfnamefont {D.}~\bibnamefont
  {Hsieh}}, \bibinfo {author} {\bibfnamefont {D.}~\bibnamefont {Qian}},
  \bibinfo {author} {\bibfnamefont {L.}~\bibnamefont {Wray}}, \bibinfo {author}
  {\bibfnamefont {Y.}~\bibnamefont {Xia}}, \bibinfo {author} {\bibfnamefont
  {Y.~S.}\ \bibnamefont {Hor}}, \bibinfo {author} {\bibfnamefont {R.~J.}\
  \bibnamefont {Cava}},\ and\ \bibinfo {author} {\bibfnamefont {M.~Z.}\
  \bibnamefont {Hasan}},\ }\href {https://doi.org/10.1038/nature06843}
  {\bibfield  {journal} {\bibinfo  {journal} {Nature}\ }\textbf {\bibinfo
  {volume} {452}},\ \bibinfo {pages} {970} (\bibinfo {year}
  {2008})}\BibitemShut {NoStop}%
\bibitem [{\citenamefont {Zhang}\ \emph {et~al.}(2009)\citenamefont {Zhang},
  \citenamefont {Liu}, \citenamefont {Qi}, \citenamefont {Dai}, \citenamefont
  {Fang},\ and\ \citenamefont {Zhang}}]{Zhang2009}%
  \BibitemOpen
  \bibfield  {author} {\bibinfo {author} {\bibfnamefont {H.}~\bibnamefont
  {Zhang}}, \bibinfo {author} {\bibfnamefont {C.-X.}\ \bibnamefont {Liu}},
  \bibinfo {author} {\bibfnamefont {X.-L.}\ \bibnamefont {Qi}}, \bibinfo
  {author} {\bibfnamefont {X.}~\bibnamefont {Dai}}, \bibinfo {author}
  {\bibfnamefont {Z.}~\bibnamefont {Fang}},\ and\ \bibinfo {author}
  {\bibfnamefont {S.-C.}\ \bibnamefont {Zhang}},\ }\href
  {https://doi.org/10.1038/nphys1270} {\bibfield  {journal} {\bibinfo
  {journal} {Nat. Phys.}\ }\textbf {\bibinfo {volume} {5}},\ \bibinfo {pages}
  {438} (\bibinfo {year} {2009})}\BibitemShut {NoStop}%
\bibitem [{\citenamefont {Xia}\ \emph {et~al.}(2009)\citenamefont {Xia},
  \citenamefont {Qian}, \citenamefont {Hsieh}, \citenamefont {Wray},
  \citenamefont {Pal}, \citenamefont {Lin}, \citenamefont {Bansil},
  \citenamefont {Grauer}, \citenamefont {Hor}, \citenamefont {Cava},\ and\
  \citenamefont {Hasan}}]{Xia2009}%
  \BibitemOpen
  \bibfield  {author} {\bibinfo {author} {\bibfnamefont {Y.}~\bibnamefont
  {Xia}}, \bibinfo {author} {\bibfnamefont {D.}~\bibnamefont {Qian}}, \bibinfo
  {author} {\bibfnamefont {D.}~\bibnamefont {Hsieh}}, \bibinfo {author}
  {\bibfnamefont {L.}~\bibnamefont {Wray}}, \bibinfo {author} {\bibfnamefont
  {A.}~\bibnamefont {Pal}}, \bibinfo {author} {\bibfnamefont {H.}~\bibnamefont
  {Lin}}, \bibinfo {author} {\bibfnamefont {A.}~\bibnamefont {Bansil}},
  \bibinfo {author} {\bibfnamefont {D.}~\bibnamefont {Grauer}}, \bibinfo
  {author} {\bibfnamefont {Y.~S.}\ \bibnamefont {Hor}}, \bibinfo {author}
  {\bibfnamefont {R.~J.}\ \bibnamefont {Cava}},\ and\ \bibinfo {author}
  {\bibfnamefont {M.~Z.}\ \bibnamefont {Hasan}},\ }\href
  {https://doi.org/10.1038/nphys1274} {\bibfield  {journal} {\bibinfo
  {journal} {Nat. Phys.}\ }\textbf {\bibinfo {volume} {5}},\ \bibinfo {pages}
  {398} (\bibinfo {year} {2009})}\BibitemShut {NoStop}%
\bibitem [{\citenamefont {Scanlon}\ \emph {et~al.}(2012)\citenamefont
  {Scanlon}, \citenamefont {King}, \citenamefont {Singh}, \citenamefont {de~la
  Torre}, \citenamefont {Walker}, \citenamefont {Balakrishnan}, \citenamefont
  {Baumberger},\ and\ \citenamefont {Catlow}}]{doi:10.1002/adma.201200187}%
  \BibitemOpen
  \bibfield  {author} {\bibinfo {author} {\bibfnamefont {D.~O.}\ \bibnamefont
  {Scanlon}}, \bibinfo {author} {\bibfnamefont {P.~D.~C.}\ \bibnamefont
  {King}}, \bibinfo {author} {\bibfnamefont {R.~P.}\ \bibnamefont {Singh}},
  \bibinfo {author} {\bibfnamefont {A.}~\bibnamefont {de~la Torre}}, \bibinfo
  {author} {\bibfnamefont {S.~M.}\ \bibnamefont {Walker}}, \bibinfo {author}
  {\bibfnamefont {G.}~\bibnamefont {Balakrishnan}}, \bibinfo {author}
  {\bibfnamefont {F.}~\bibnamefont {Baumberger}},\ and\ \bibinfo {author}
  {\bibfnamefont {C.~R.~A.}\ \bibnamefont {Catlow}},\ }\href
  {https://doi.org/10.1002/adma.201200187} {\bibfield  {journal} {\bibinfo
  {journal} {Adv. Mater.}\ }\textbf {\bibinfo {volume} {24}},\ \bibinfo {pages}
  {2154} (\bibinfo {year} {2012})}\BibitemShut {NoStop}%
\bibitem [{\citenamefont {Xu}\ \emph {et~al.}(2014)\citenamefont {Xu},
  \citenamefont {Miotkowski}, \citenamefont {Liu}, \citenamefont {Tian},
  \citenamefont {Nam}, \citenamefont {Alidoust}, \citenamefont {Hu},
  \citenamefont {Shih}, \citenamefont {Hasan},\ and\ \citenamefont
  {Chen}}]{Xu2014}%
  \BibitemOpen
  \bibfield  {author} {\bibinfo {author} {\bibfnamefont {Y.}~\bibnamefont
  {Xu}}, \bibinfo {author} {\bibfnamefont {I.}~\bibnamefont {Miotkowski}},
  \bibinfo {author} {\bibfnamefont {C.}~\bibnamefont {Liu}}, \bibinfo {author}
  {\bibfnamefont {J.}~\bibnamefont {Tian}}, \bibinfo {author} {\bibfnamefont
  {H.}~\bibnamefont {Nam}}, \bibinfo {author} {\bibfnamefont {N.}~\bibnamefont
  {Alidoust}}, \bibinfo {author} {\bibfnamefont {J.}~\bibnamefont {Hu}},
  \bibinfo {author} {\bibfnamefont {C.-K.}\ \bibnamefont {Shih}}, \bibinfo
  {author} {\bibfnamefont {M.~Z.}\ \bibnamefont {Hasan}},\ and\ \bibinfo
  {author} {\bibfnamefont {Y.~P.}\ \bibnamefont {Chen}},\ }\href
  {https://doi.org/10.1038/nphys3140} {\bibfield  {journal} {\bibinfo
  {journal} {Nat. Phys.}\ }\textbf {\bibinfo {volume} {10}},\ \bibinfo {pages}
  {956} (\bibinfo {year} {2014})}\BibitemShut {NoStop}%
\bibitem [{\citenamefont {Ginley}\ \emph {et~al.}(2016)\citenamefont {Ginley},
  \citenamefont {Wang},\ and\ \citenamefont {Law}}]{Ginley_2016}%
  \BibitemOpen
  \bibfield  {author} {\bibinfo {author} {\bibfnamefont {T.}~\bibnamefont
  {Ginley}}, \bibinfo {author} {\bibfnamefont {Y.}~\bibnamefont {Wang}},\ and\
  \bibinfo {author} {\bibfnamefont {S.}~\bibnamefont {Law}},\ }\href
  {https://doi.org/10.3390/cryst6110154} {\bibfield  {journal} {\bibinfo
  {journal} {Crystals}\ }\textbf {\bibinfo {volume} {6}},\ \bibinfo {pages}
  {154} (\bibinfo {year} {2016})}\BibitemShut {NoStop}%
\bibitem [{\citenamefont {Mussler}(2020)}]{doi:10.1002/pssb.202000007}%
  \BibitemOpen
  \bibfield  {author} {\bibinfo {author} {\bibfnamefont {G.}~\bibnamefont
  {Mussler}},\ }\href {https://doi.org/10.1002/pssb.202000007} {\bibfield
  {journal} {\bibinfo  {journal} {Phys. Status Solidi B}\ }\textbf {\bibinfo
  {volume} {257}},\ \bibinfo {pages} {2000007} (\bibinfo {year}
  {2020})}\BibitemShut {NoStop}%
\bibitem [{\citenamefont {Chen}\ \emph {et~al.}(2009)\citenamefont {Chen},
  \citenamefont {Analytis}, \citenamefont {Chu}, \citenamefont {Liu},
  \citenamefont {Mo}, \citenamefont {Qi}, \citenamefont {Zhang}, \citenamefont
  {Lu}, \citenamefont {Dai}, \citenamefont {Fang}, \citenamefont {Zhang},
  \citenamefont {Fisher}, \citenamefont {Hussain},\ and\ \citenamefont
  {Shen}}]{Chen178}%
  \BibitemOpen
  \bibfield  {author} {\bibinfo {author} {\bibfnamefont {Y.~L.}\ \bibnamefont
  {Chen}}, \bibinfo {author} {\bibfnamefont {J.~G.}\ \bibnamefont {Analytis}},
  \bibinfo {author} {\bibfnamefont {J.-H.}\ \bibnamefont {Chu}}, \bibinfo
  {author} {\bibfnamefont {Z.~K.}\ \bibnamefont {Liu}}, \bibinfo {author}
  {\bibfnamefont {S.-K.}\ \bibnamefont {Mo}}, \bibinfo {author} {\bibfnamefont
  {X.~L.}\ \bibnamefont {Qi}}, \bibinfo {author} {\bibfnamefont {H.~J.}\
  \bibnamefont {Zhang}}, \bibinfo {author} {\bibfnamefont {D.~H.}\ \bibnamefont
  {Lu}}, \bibinfo {author} {\bibfnamefont {X.}~\bibnamefont {Dai}}, \bibinfo
  {author} {\bibfnamefont {Z.}~\bibnamefont {Fang}}, \bibinfo {author}
  {\bibfnamefont {S.~C.}\ \bibnamefont {Zhang}}, \bibinfo {author}
  {\bibfnamefont {I.~R.}\ \bibnamefont {Fisher}}, \bibinfo {author}
  {\bibfnamefont {Z.}~\bibnamefont {Hussain}},\ and\ \bibinfo {author}
  {\bibfnamefont {Z.-X.}\ \bibnamefont {Shen}},\ }\href
  {https://doi.org/10.1126/science.1173034} {\bibfield  {journal} {\bibinfo
  {journal} {Science}\ }\textbf {\bibinfo {volume} {325}},\ \bibinfo {pages}
  {178} (\bibinfo {year} {2009})}\BibitemShut {NoStop}%
\bibitem [{\citenamefont {Rüßmann}\ \emph {et~al.}(2019)\citenamefont
  {Rüßmann}, \citenamefont {Mavropoulos},\ and\ \citenamefont
  {Blügel}}]{RUMANN2019258}%
  \BibitemOpen
  \bibfield  {author} {\bibinfo {author} {\bibfnamefont {P.}~\bibnamefont
  {Rüßmann}}, \bibinfo {author} {\bibfnamefont {P.}~\bibnamefont
  {Mavropoulos}},\ and\ \bibinfo {author} {\bibfnamefont {S.}~\bibnamefont
  {Blügel}},\ }\href
  {https://doi.org/https://doi.org/10.1016/j.jpcs.2017.12.009} {\bibfield
  {journal} {\bibinfo  {journal} {J. phys. Chem. Solids}\ }\textbf {\bibinfo
  {volume} {128}},\ \bibinfo {pages} {258 } (\bibinfo {year}
  {2019})}\BibitemShut {NoStop}%
\bibitem [{\citenamefont {Ren}\ \emph {et~al.}(2011)\citenamefont {Ren},
  \citenamefont {Taskin}, \citenamefont {Sasaki}, \citenamefont {Segawa},\ and\
  \citenamefont {Ando}}]{PhysRevB.84.165311}%
  \BibitemOpen
  \bibfield  {author} {\bibinfo {author} {\bibfnamefont {Z.}~\bibnamefont
  {Ren}}, \bibinfo {author} {\bibfnamefont {A.~A.}\ \bibnamefont {Taskin}},
  \bibinfo {author} {\bibfnamefont {S.}~\bibnamefont {Sasaki}}, \bibinfo
  {author} {\bibfnamefont {K.}~\bibnamefont {Segawa}},\ and\ \bibinfo {author}
  {\bibfnamefont {Y.}~\bibnamefont {Ando}},\ }\href
  {https://doi.org/10.1103/PhysRevB.84.165311} {\bibfield  {journal} {\bibinfo
  {journal} {Phys. Rev. B}\ }\textbf {\bibinfo {volume} {84}},\ \bibinfo
  {pages} {165311} (\bibinfo {year} {2011})}\BibitemShut {NoStop}%
\bibitem [{\citenamefont {Arakane}\ \emph {et~al.}(2012)\citenamefont
  {Arakane}, \citenamefont {Sato}, \citenamefont {Souma}, \citenamefont
  {Kosaka}, \citenamefont {Nakayama}, \citenamefont {Komatsu}, \citenamefont
  {Takahashi}, \citenamefont {Ren}, \citenamefont {Segawa},\ and\ \citenamefont
  {Ando}}]{Arakane2012}%
  \BibitemOpen
  \bibfield  {author} {\bibinfo {author} {\bibfnamefont {T.}~\bibnamefont
  {Arakane}}, \bibinfo {author} {\bibfnamefont {T.}~\bibnamefont {Sato}},
  \bibinfo {author} {\bibfnamefont {S.}~\bibnamefont {Souma}}, \bibinfo
  {author} {\bibfnamefont {K.}~\bibnamefont {Kosaka}}, \bibinfo {author}
  {\bibfnamefont {K.}~\bibnamefont {Nakayama}}, \bibinfo {author}
  {\bibfnamefont {M.}~\bibnamefont {Komatsu}}, \bibinfo {author} {\bibfnamefont
  {T.}~\bibnamefont {Takahashi}}, \bibinfo {author} {\bibfnamefont
  {Z.}~\bibnamefont {Ren}}, \bibinfo {author} {\bibfnamefont {K.}~\bibnamefont
  {Segawa}},\ and\ \bibinfo {author} {\bibfnamefont {Y.}~\bibnamefont {Ando}},\
  }\href {https://doi.org/10.1038/ncomms1639} {\bibfield  {journal} {\bibinfo
  {journal} {Nat. Commun.}\ }\textbf {\bibinfo {volume} {3}},\ \bibinfo {pages}
  {636} (\bibinfo {year} {2012})}\BibitemShut {NoStop}%
\bibitem [{\citenamefont {Xia}\ \emph {et~al.}(2013)\citenamefont {Xia},
  \citenamefont {Ren}, \citenamefont {Sulaev}, \citenamefont {Liu},
  \citenamefont {Shen},\ and\ \citenamefont {Wang}}]{PhysRevB.87.085442}%
  \BibitemOpen
  \bibfield  {author} {\bibinfo {author} {\bibfnamefont {B.}~\bibnamefont
  {Xia}}, \bibinfo {author} {\bibfnamefont {P.}~\bibnamefont {Ren}}, \bibinfo
  {author} {\bibfnamefont {A.}~\bibnamefont {Sulaev}}, \bibinfo {author}
  {\bibfnamefont {P.}~\bibnamefont {Liu}}, \bibinfo {author} {\bibfnamefont
  {S.-Q.}\ \bibnamefont {Shen}},\ and\ \bibinfo {author} {\bibfnamefont
  {L.}~\bibnamefont {Wang}},\ }\href
  {https://doi.org/10.1103/PhysRevB.87.085442} {\bibfield  {journal} {\bibinfo
  {journal} {Phys. Rev. B}\ }\textbf {\bibinfo {volume} {87}},\ \bibinfo
  {pages} {085442} (\bibinfo {year} {2013})}\BibitemShut {NoStop}%
\bibitem [{\citenamefont {Borgwardt}\ \emph {et~al.}(2016)\citenamefont
  {Borgwardt}, \citenamefont {Lux}, \citenamefont {Vergara}, \citenamefont
  {Wang}, \citenamefont {Taskin}, \citenamefont {Segawa}, \citenamefont {van
  Loosdrecht}, \citenamefont {Ando}, \citenamefont {Rosch},\ and\ \citenamefont
  {Gr\"uninger}}]{PhysRevB.93.245149}%
  \BibitemOpen
  \bibfield  {author} {\bibinfo {author} {\bibfnamefont {N.}~\bibnamefont
  {Borgwardt}}, \bibinfo {author} {\bibfnamefont {J.}~\bibnamefont {Lux}},
  \bibinfo {author} {\bibfnamefont {I.}~\bibnamefont {Vergara}}, \bibinfo
  {author} {\bibfnamefont {Z.}~\bibnamefont {Wang}}, \bibinfo {author}
  {\bibfnamefont {A.~A.}\ \bibnamefont {Taskin}}, \bibinfo {author}
  {\bibfnamefont {K.}~\bibnamefont {Segawa}}, \bibinfo {author} {\bibfnamefont
  {P.~H.~M.}\ \bibnamefont {van Loosdrecht}}, \bibinfo {author} {\bibfnamefont
  {Y.}~\bibnamefont {Ando}}, \bibinfo {author} {\bibfnamefont {A.}~\bibnamefont
  {Rosch}},\ and\ \bibinfo {author} {\bibfnamefont {M.}~\bibnamefont
  {Gr\"uninger}},\ }\href {https://doi.org/10.1103/PhysRevB.93.245149}
  {\bibfield  {journal} {\bibinfo  {journal} {Phys. Rev. B}\ }\textbf {\bibinfo
  {volume} {93}},\ \bibinfo {pages} {245149} (\bibinfo {year}
  {2016})}\BibitemShut {NoStop}%
\bibitem [{\citenamefont {Zhang}\ \emph {et~al.}(2013)\citenamefont {Zhang},
  \citenamefont {Feng}, \citenamefont {Guo}, \citenamefont {Ou}, \citenamefont
  {Zhang}, \citenamefont {Li}, \citenamefont {Wang}, \citenamefont {Chen},
  \citenamefont {Xue}, \citenamefont {Ma}, \citenamefont {He},\ and\
  \citenamefont {Wang}}]{doi:10.1002/pssr.201206391}%
  \BibitemOpen
  \bibfield  {author} {\bibinfo {author} {\bibfnamefont {Z.}~\bibnamefont
  {Zhang}}, \bibinfo {author} {\bibfnamefont {X.}~\bibnamefont {Feng}},
  \bibinfo {author} {\bibfnamefont {M.}~\bibnamefont {Guo}}, \bibinfo {author}
  {\bibfnamefont {Y.}~\bibnamefont {Ou}}, \bibinfo {author} {\bibfnamefont
  {J.}~\bibnamefont {Zhang}}, \bibinfo {author} {\bibfnamefont
  {K.}~\bibnamefont {Li}}, \bibinfo {author} {\bibfnamefont {L.}~\bibnamefont
  {Wang}}, \bibinfo {author} {\bibfnamefont {X.}~\bibnamefont {Chen}}, \bibinfo
  {author} {\bibfnamefont {Q.}~\bibnamefont {Xue}}, \bibinfo {author}
  {\bibfnamefont {X.}~\bibnamefont {Ma}}, \bibinfo {author} {\bibfnamefont
  {K.}~\bibnamefont {He}},\ and\ \bibinfo {author} {\bibfnamefont
  {Y.}~\bibnamefont {Wang}},\ }\href {https://doi.org/10.1002/pssr.201206391}
  {\bibfield  {journal} {\bibinfo  {journal} {Phys. Status Solidi-R}\ }\textbf
  {\bibinfo {volume} {7}},\ \bibinfo {pages} {142} (\bibinfo {year}
  {2013})}\BibitemShut {NoStop}%
\bibitem [{\citenamefont {Eschbach}\ \emph {et~al.}(2015)\citenamefont
  {Eschbach}, \citenamefont {M{\l}y{\'{n}}czak}, \citenamefont {Kellner},
  \citenamefont {Kampmeier}, \citenamefont {Lanius}, \citenamefont {Neumann},
  \citenamefont {Weyrich}, \citenamefont {Gehlmann}, \citenamefont
  {Gospodari{\v{c}}}, \citenamefont {D{\"o}ring}, \citenamefont {Mussler},
  \citenamefont {Demarina}, \citenamefont {Luysberg}, \citenamefont
  {Bihlmayer}, \citenamefont {Sch{\"a}pers}, \citenamefont {Plucinski},
  \citenamefont {Bl{\"u}gel}, \citenamefont {Morgenstern}, \citenamefont
  {Schneider},\ and\ \citenamefont {Gr{\"u}tzmacher}}]{Eschbach2015}%
  \BibitemOpen
  \bibfield  {author} {\bibinfo {author} {\bibfnamefont {M.}~\bibnamefont
  {Eschbach}}, \bibinfo {author} {\bibfnamefont {E.}~\bibnamefont
  {M{\l}y{\'{n}}czak}}, \bibinfo {author} {\bibfnamefont {J.}~\bibnamefont
  {Kellner}}, \bibinfo {author} {\bibfnamefont {J.}~\bibnamefont {Kampmeier}},
  \bibinfo {author} {\bibfnamefont {M.}~\bibnamefont {Lanius}}, \bibinfo
  {author} {\bibfnamefont {E.}~\bibnamefont {Neumann}}, \bibinfo {author}
  {\bibfnamefont {C.}~\bibnamefont {Weyrich}}, \bibinfo {author} {\bibfnamefont
  {M.}~\bibnamefont {Gehlmann}}, \bibinfo {author} {\bibfnamefont
  {P.}~\bibnamefont {Gospodari{\v{c}}}}, \bibinfo {author} {\bibfnamefont
  {S.}~\bibnamefont {D{\"o}ring}}, \bibinfo {author} {\bibfnamefont
  {G.}~\bibnamefont {Mussler}}, \bibinfo {author} {\bibfnamefont
  {N.}~\bibnamefont {Demarina}}, \bibinfo {author} {\bibfnamefont
  {M.}~\bibnamefont {Luysberg}}, \bibinfo {author} {\bibfnamefont
  {G.}~\bibnamefont {Bihlmayer}}, \bibinfo {author} {\bibfnamefont
  {T.}~\bibnamefont {Sch{\"a}pers}}, \bibinfo {author} {\bibfnamefont
  {L.}~\bibnamefont {Plucinski}}, \bibinfo {author} {\bibfnamefont
  {S.}~\bibnamefont {Bl{\"u}gel}}, \bibinfo {author} {\bibfnamefont
  {M.}~\bibnamefont {Morgenstern}}, \bibinfo {author} {\bibfnamefont {C.~M.}\
  \bibnamefont {Schneider}},\ and\ \bibinfo {author} {\bibfnamefont
  {D.}~\bibnamefont {Gr{\"u}tzmacher}},\ }\href
  {https://doi.org/10.1038/ncomms9816} {\bibfield  {journal} {\bibinfo
  {journal} {Nat. Commun.}\ }\textbf {\bibinfo {volume} {6}},\ \bibinfo {pages}
  {8816} (\bibinfo {year} {2015})}\BibitemShut {NoStop}%
\bibitem [{\citenamefont {Backes}\ \emph {et~al.}(2017)\citenamefont {Backes},
  \citenamefont {Huang}, \citenamefont {Mansell}, \citenamefont {Lanius},
  \citenamefont {Kampmeier}, \citenamefont {Ritchie}, \citenamefont {Mussler},
  \citenamefont {Gumbs}, \citenamefont {Gr\"utzmacher},\ and\ \citenamefont
  {Narayan}}]{PhysRevB.96.125125}%
  \BibitemOpen
  \bibfield  {author} {\bibinfo {author} {\bibfnamefont {D.}~\bibnamefont
  {Backes}}, \bibinfo {author} {\bibfnamefont {D.}~\bibnamefont {Huang}},
  \bibinfo {author} {\bibfnamefont {R.}~\bibnamefont {Mansell}}, \bibinfo
  {author} {\bibfnamefont {M.}~\bibnamefont {Lanius}}, \bibinfo {author}
  {\bibfnamefont {J.}~\bibnamefont {Kampmeier}}, \bibinfo {author}
  {\bibfnamefont {D.}~\bibnamefont {Ritchie}}, \bibinfo {author} {\bibfnamefont
  {G.}~\bibnamefont {Mussler}}, \bibinfo {author} {\bibfnamefont
  {G.}~\bibnamefont {Gumbs}}, \bibinfo {author} {\bibfnamefont
  {D.}~\bibnamefont {Gr\"utzmacher}},\ and\ \bibinfo {author} {\bibfnamefont
  {V.}~\bibnamefont {Narayan}},\ }\href
  {https://doi.org/10.1103/PhysRevB.96.125125} {\bibfield  {journal} {\bibinfo
  {journal} {Phys. Rev. B}\ }\textbf {\bibinfo {volume} {96}},\ \bibinfo
  {pages} {125125} (\bibinfo {year} {2017})}\BibitemShut {NoStop}%
\bibitem [{\citenamefont {Gao}\ \emph {et~al.}(2012)\citenamefont {Gao},
  \citenamefont {Gehring}, \citenamefont {Burghard},\ and\ \citenamefont
  {Kern}}]{Gao2012}%
  \BibitemOpen
  \bibfield  {author} {\bibinfo {author} {\bibfnamefont {B.~F.}\ \bibnamefont
  {Gao}}, \bibinfo {author} {\bibfnamefont {P.}~\bibnamefont {Gehring}},
  \bibinfo {author} {\bibfnamefont {M.}~\bibnamefont {Burghard}},\ and\
  \bibinfo {author} {\bibfnamefont {K.}~\bibnamefont {Kern}},\ }\href
  {https://doi.org/10.1063/1.4719196} {\bibfield  {journal} {\bibinfo
  {journal} {Appl. Phys. Lett.}\ }\textbf {\bibinfo {volume} {100}},\ \bibinfo
  {pages} {212402} (\bibinfo {year} {2012})}\BibitemShut {NoStop}%
\bibitem [{sup()}]{supplemental}%
  \BibitemOpen
  \href@noop {} {\bibinfo {title} {{ See Supplemental Material at [ ] for
  resistivity values ver-sus sample thickness and Hall
  measurements.}}}\BibitemShut {Stop}%
\bibitem [{Note1()}]{Note1}%
  \BibitemOpen
  \bibinfo {note} {Within each series, the BS thickness remains constant while
  the BSTS thickness gradually increases. The limit of $t\rightarrow \infty $
  therefore means that the BSTS thickness approaches infinity: any thickness
  independent contribution to conduction like the TSS or any contribution from
  the BS vanishes.}\BibitemShut {Stop}%
\bibitem [{\citenamefont {Skinner}\ \emph {et~al.}(2012)\citenamefont
  {Skinner}, \citenamefont {Chen},\ and\ \citenamefont
  {Shklovskii}}]{PhysRevLett.109.176801}%
  \BibitemOpen
  \bibfield  {author} {\bibinfo {author} {\bibfnamefont {B.}~\bibnamefont
  {Skinner}}, \bibinfo {author} {\bibfnamefont {T.}~\bibnamefont {Chen}},\ and\
  \bibinfo {author} {\bibfnamefont {B.~I.}\ \bibnamefont {Shklovskii}},\ }\href
  {https://doi.org/10.1103/PhysRevLett.109.176801} {\bibfield  {journal}
  {\bibinfo  {journal} {Phys. Rev. Lett.}\ }\textbf {\bibinfo {volume} {109}},\
  \bibinfo {pages} {176801} (\bibinfo {year} {2012})}\BibitemShut {NoStop}%
\bibitem [{\citenamefont {Skinner}\ \emph {et~al.}(2013)\citenamefont
  {Skinner}, \citenamefont {Chen},\ and\ \citenamefont
  {Shklovskii}}]{Skinner2013}%
  \BibitemOpen
  \bibfield  {author} {\bibinfo {author} {\bibfnamefont {B.}~\bibnamefont
  {Skinner}}, \bibinfo {author} {\bibfnamefont {T.}~\bibnamefont {Chen}},\ and\
  \bibinfo {author} {\bibfnamefont {B.~I.}\ \bibnamefont {Shklovskii}},\ }\href
  {https://doi.org/10.1134/S1063776113110150} {\bibfield  {journal} {\bibinfo
  {journal} {J. Exp. Theor. Phys.+}\ }\textbf {\bibinfo {volume} {117}},\
  \bibinfo {pages} {579} (\bibinfo {year} {2013})}\BibitemShut {NoStop}%
\bibitem [{\citenamefont {Chen}\ and\ \citenamefont
  {Shklovskii}(2013)}]{PhysRevB.87.165119}%
  \BibitemOpen
  \bibfield  {author} {\bibinfo {author} {\bibfnamefont {T.}~\bibnamefont
  {Chen}}\ and\ \bibinfo {author} {\bibfnamefont {B.~I.}\ \bibnamefont
  {Shklovskii}},\ }\href {https://doi.org/10.1103/PhysRevB.87.165119}
  {\bibfield  {journal} {\bibinfo  {journal} {Phys. Rev. B}\ }\textbf {\bibinfo
  {volume} {87}},\ \bibinfo {pages} {165119} (\bibinfo {year}
  {2013})}\BibitemShut {NoStop}%
\bibitem [{\citenamefont {Rischau}\ \emph {et~al.}(2016)\citenamefont
  {Rischau}, \citenamefont {Ubaldini}, \citenamefont {Giannini},\ and\
  \citenamefont {van~der Beek}}]{Rischau_2016}%
  \BibitemOpen
  \bibfield  {author} {\bibinfo {author} {\bibfnamefont {C.~W.}\ \bibnamefont
  {Rischau}}, \bibinfo {author} {\bibfnamefont {A.}~\bibnamefont {Ubaldini}},
  \bibinfo {author} {\bibfnamefont {E.}~\bibnamefont {Giannini}},\ and\
  \bibinfo {author} {\bibfnamefont {C.~J.}\ \bibnamefont {van~der Beek}},\
  }\href {https://doi.org/10.1088/1367-2630/18/7/073024} {\bibfield  {journal}
  {\bibinfo  {journal} {New J. Phys.}\ }\textbf {\bibinfo {volume} {18}},\
  \bibinfo {pages} {073024} (\bibinfo {year} {2016})}\BibitemShut {NoStop}%
\bibitem [{\citenamefont {Lu}\ and\ \citenamefont
  {Shen}(2014)}]{10.1117/12.2063426}%
  \BibitemOpen
  \bibfield  {author} {\bibinfo {author} {\bibfnamefont {H.-Z.}\ \bibnamefont
  {Lu}}\ and\ \bibinfo {author} {\bibfnamefont {S.-Q.}\ \bibnamefont {Shen}},\
  }\href {https://doi.org/10.1117/12.2063426} {\bibfield  {journal} {\bibinfo
  {journal} {Proc. SPIE Spintronics VII}\ }\textbf {\bibinfo {volume} {9167}},\
  \bibinfo {pages} {263 } (\bibinfo {year} {2014})}\BibitemShut {NoStop}%
\bibitem [{\citenamefont {Garate}\ and\ \citenamefont
  {Glazman}(2012)}]{PhysRevB.86.035422}%
  \BibitemOpen
  \bibfield  {author} {\bibinfo {author} {\bibfnamefont {I.}~\bibnamefont
  {Garate}}\ and\ \bibinfo {author} {\bibfnamefont {L.}~\bibnamefont
  {Glazman}},\ }\href {https://doi.org/10.1103/PhysRevB.86.035422} {\bibfield
  {journal} {\bibinfo  {journal} {Phys. Rev. B}\ }\textbf {\bibinfo {volume}
  {86}},\ \bibinfo {pages} {035422} (\bibinfo {year} {2012})}\BibitemShut
  {NoStop}%
\bibitem [{\citenamefont {Abrikosov}(1998)}]{PhysRevB.58.2788}%
  \BibitemOpen
  \bibfield  {author} {\bibinfo {author} {\bibfnamefont {A.~A.}\ \bibnamefont
  {Abrikosov}},\ }\href {https://doi.org/10.1103/PhysRevB.58.2788} {\bibfield
  {journal} {\bibinfo  {journal} {Phys. Rev. B}\ }\textbf {\bibinfo {volume}
  {58}},\ \bibinfo {pages} {2788} (\bibinfo {year} {1998})}\BibitemShut
  {NoStop}%
\bibitem [{\citenamefont {Parish}\ and\ \citenamefont
  {Littlewood}(2003)}]{Parish2003}%
  \BibitemOpen
  \bibfield  {author} {\bibinfo {author} {\bibfnamefont {M.~M.}\ \bibnamefont
  {Parish}}\ and\ \bibinfo {author} {\bibfnamefont {P.~B.}\ \bibnamefont
  {Littlewood}},\ }\href {https://doi.org/10.1038/nature02073} {\bibfield
  {journal} {\bibinfo  {journal} {Nature}\ }\textbf {\bibinfo {volume} {426}},\
  \bibinfo {pages} {162} (\bibinfo {year} {2003})}\BibitemShut {NoStop}%
\bibitem [{\citenamefont {Parish}\ and\ \citenamefont
  {Littlewood}(2005)}]{PhysRevB.72.094417}%
  \BibitemOpen
  \bibfield  {author} {\bibinfo {author} {\bibfnamefont {M.~M.}\ \bibnamefont
  {Parish}}\ and\ \bibinfo {author} {\bibfnamefont {P.~B.}\ \bibnamefont
  {Littlewood}},\ }\href {https://doi.org/10.1103/PhysRevB.72.094417}
  {\bibfield  {journal} {\bibinfo  {journal} {Phys. Rev. B}\ }\textbf {\bibinfo
  {volume} {72}},\ \bibinfo {pages} {094417} (\bibinfo {year}
  {2005})}\BibitemShut {NoStop}%
\bibitem [{\citenamefont {Hikami}\ \emph {et~al.}(1980)\citenamefont {Hikami},
  \citenamefont {Larkin},\ and\ \citenamefont {Nagaoka}}]{10.1143/PTP.63.707}%
  \BibitemOpen
  \bibfield  {author} {\bibinfo {author} {\bibfnamefont {S.}~\bibnamefont
  {Hikami}}, \bibinfo {author} {\bibfnamefont {A.~I.}\ \bibnamefont {Larkin}},\
  and\ \bibinfo {author} {\bibfnamefont {Y.}~\bibnamefont {Nagaoka}},\ }\href
  {https://doi.org/10.1143/PTP.63.707} {\bibfield  {journal} {\bibinfo
  {journal} {Prog. Theor. Phys.}\ }\textbf {\bibinfo {volume} {63}},\ \bibinfo
  {pages} {707} (\bibinfo {year} {1980})}\BibitemShut {NoStop}%
\bibitem [{\citenamefont {Brahlek}\ \emph {et~al.}(2014)\citenamefont
  {Brahlek}, \citenamefont {Koirala}, \citenamefont {Salehi}, \citenamefont
  {Bansal},\ and\ \citenamefont {Oh}}]{PhysRevLett.113.026801}%
  \BibitemOpen
  \bibfield  {author} {\bibinfo {author} {\bibfnamefont {M.}~\bibnamefont
  {Brahlek}}, \bibinfo {author} {\bibfnamefont {N.}~\bibnamefont {Koirala}},
  \bibinfo {author} {\bibfnamefont {M.}~\bibnamefont {Salehi}}, \bibinfo
  {author} {\bibfnamefont {N.}~\bibnamefont {Bansal}},\ and\ \bibinfo {author}
  {\bibfnamefont {S.}~\bibnamefont {Oh}},\ }\href
  {https://doi.org/10.1103/PhysRevLett.113.026801} {\bibfield  {journal}
  {\bibinfo  {journal} {Phys. Rev. Lett.}\ }\textbf {\bibinfo {volume} {113}},\
  \bibinfo {pages} {026801} (\bibinfo {year} {2014})}\BibitemShut {NoStop}%
\bibitem [{\citenamefont {Wang}\ \emph {et~al.}(2016)\citenamefont {Wang},
  \citenamefont {Gao},\ and\ \citenamefont {Li}}]{Wang2016}%
  \BibitemOpen
  \bibfield  {author} {\bibinfo {author} {\bibfnamefont {W.~J.}\ \bibnamefont
  {Wang}}, \bibinfo {author} {\bibfnamefont {K.~H.}\ \bibnamefont {Gao}},\ and\
  \bibinfo {author} {\bibfnamefont {Z.~Q.}\ \bibnamefont {Li}},\ }\href
  {https://doi.org/10.1038/srep25291} {\bibfield  {journal} {\bibinfo
  {journal} {Sci. Rep.}\ }\textbf {\bibinfo {volume} {6}},\ \bibinfo {pages}
  {25291} (\bibinfo {year} {2016})}\BibitemShut {NoStop}%
\bibitem [{\citenamefont {Fleury}\ and\ \citenamefont
  {Worlock}(1968)}]{PhysRev.174.613}%
  \BibitemOpen
  \bibfield  {author} {\bibinfo {author} {\bibfnamefont {P.~A.}\ \bibnamefont
  {Fleury}}\ and\ \bibinfo {author} {\bibfnamefont {J.~M.}\ \bibnamefont
  {Worlock}},\ }\href {https://doi.org/10.1103/PhysRev.174.613} {\bibfield
  {journal} {\bibinfo  {journal} {Phys. Rev.}\ }\textbf {\bibinfo {volume}
  {174}},\ \bibinfo {pages} {613} (\bibinfo {year} {1968})}\BibitemShut
  {NoStop}%
\bibitem [{\citenamefont {M\"uller}\ and\ \citenamefont
  {Burkard}(1979)}]{PhysRevB.19.3593}%
  \BibitemOpen
  \bibfield  {author} {\bibinfo {author} {\bibfnamefont {K.~A.}\ \bibnamefont
  {M\"uller}}\ and\ \bibinfo {author} {\bibfnamefont {H.}~\bibnamefont
  {Burkard}},\ }\href {https://doi.org/10.1103/PhysRevB.19.3593} {\bibfield
  {journal} {\bibinfo  {journal} {Phys. Rev. B}\ }\textbf {\bibinfo {volume}
  {19}},\ \bibinfo {pages} {3593} (\bibinfo {year} {1979})}\BibitemShut
  {NoStop}%
\bibitem [{Note2()}]{Note2}%
  \BibitemOpen
  \bibinfo {note} {To account for hysteresis effects, the front-gate was swept
  at 4.2$\protect \,$K from +30$\protect \,$V to -30$\protect \,$V twice before
  the measurement was started. Then, the back-gate was set in steps of
  1.5$\protect \,$V in positive direction. After 10 minutes, to prevent
  time-dependent influences, the front-gate was swept from +30$\protect \,$V to
  -30$\protect \,$V with 60$\protect \,$mV/s. After the measurement, the sample
  was brought to room-temperature to reset both gates and the procedure was
  repeated with the back-gate being set into negative direction. Due to this
  required reset of the gates a small discrepancy between the measurements for
  positive and negative back-gate direction is unavoidable. To account for
  this, the $R_\protect \mathrm {S}$ values for positive back-gate voltage were
  shifted by 3.3$\protect \,$V front-gate in Fig. 5d).}\BibitemShut {Stop}%
\end{thebibliography}%

\end{document}